\documentclass[a4paper,preprint,unsortedaddress,onecolumn]{revtex4-1}

\usepackage[fleqn]{amsmath}
\usepackage{amssymb}
\usepackage[dvips]{graphicx}
\usepackage{color}
\usepackage{tabularx}
\usepackage{algorithm}
\usepackage{algorithmic}


\begin{document}

\title{Formulation of Liouville's Theorem for Grand Ensemble Molecular Simulations}
\author{Luigi Delle Site}
\affiliation{Institute for Mathematics, Freie Universit\"at Berlin, Germany}
\email{luigi.dellesite@fu-berlin.de}
\begin{abstract}
Liouville's theorem in a grand ensemble, that is for situations where a system is in equilibrium with a reservoir of energy and particles, is a {subject} that, so far, has not been explicitly treated {in literature related to molecular simulation}. Instead Liouville's theorem, {a central concept for the correct employment of molecular simulation techniques,} is implicitly considered only within the framework of systems where the total number of particles is fixed. However the pressing demand of applied science in treating open systems leads to the question of the existence and possible exact formulation of Liouville's theorem when the number of particles changes during the dynamical evolution of the system. The intention of this note is to stimulate a debate about this crucial issue for molecular simulation.\\

{\bf PACS code: 05.20.Gg 02.70.-c 05.20.Jj}
\end{abstract}
\maketitle

\section{Introduction}
We propose a problem that, in our knowledge {and at least in the field of molecular simulation}, has not been explicitly treated before, namely whether or not it is possible a rigorous formulation of Liouville's theorem (and corresponding operator) when a system is characterized by a varying number of particles. In the following discussion we will restrict the treatment to classical systems {due to the direct implications for classical molecular simulations}. {Actually, as it will be discussed later, there exists a rich literature for quantum open systems whose formal results can be employed to define, in a rather precise way, the specific concepts needed in classical molecular simulation.}
{We will start from the very general concept of Lindblad operator \cite{linda} for open quantum systems and consider the (simpler) sub-case of a classical system. The main result which emerges from this analysis is the central role played by the (formal) definition of the reservoir, implicitly encrypted into the definition of Lindblad operator. We will consider one, physically well founded, definition of reservoir, the so-called Bergmann-Lebowitz model \cite{leb1,leb2,leb3}, and analyze the consequences when its formal concepts are translated into practical definitions for numerical calculations in a molecular simulation framework.  It is important to notice that the Bergmann-Lebowitz model have been already applied in molecular dynamics studies and led to satisfactory results \cite{njp,pi1}; thus its  positive application rises the need of a mathematical and physical analysis. In particular we will discuss the existence and meaning of Liouville's operator and Liouville's theorem for systems with varying $N$. These two concepts, in case of fixed $N$, are directly used in the calculation of key physical quantities}, thus it is of interest to understand what happens when $N$ is variable {(for a basic theoretical formulation of the different aspects of the problem see also the summary reported in Ref.\cite{penrose} and references therein)}. In fact, Liouville's theorem is central for  the correct {\bf physical} definition of quantities calculated via ensemble averaging (that is the main aim of molecular simulation)\cite{tuck}; statistical time correlation functions are relevant examples {where, in particular, the definition of Liouville's operator is explicitly needed (see discussion later). In fact, we will see that when such quantities are calculated in an open system, they require a technical redefinition (directly linked to the definition of Lindblad operator) and a careful reinterpretation of their physical meaning in terms of a relation between the locality in space and locality in time \cite{njp}.} 
Other models, in Molecular Dynamics, are based on the unphysical assumption that the number of particles $N$ is considered as a continuous variable {; also in this case numerical results are satisfactory \cite{flor}, but conceptually the idea is not consistent with first principles of open systems as derived in Ref.\cite{open}}.
{The relevant aspect of this discussion is that in molecular dynamics the existence of a first principle behind the definition of certain physical quantities is often not necessary (from the technical point of view) for their numerical calculation}. The consequence is that practical definitions (and corresponding calculation procedures) are empirical, however their transferability to other situations requires the existence of a physical well posed principle. For example, {consistently with the discuss above, in Refs.\cite{noi,kk-prot}, equilibrium time correlation functions for open boundary systems are calculated on the basis of physical intuition but without explicitly specifying what is the Liouville operator {of the atomistic region considered}. {It is assumed that {such operator} exists and it is a straightforward extension of the case with fixed $N$ {(see also note \cite{noteaddress} for more details)}. Of course for the systems considered in Refs.\cite{noi,kk-prot} the results can be usefully employed for the specific purposes of the study.} However, in general, the question is not so trivial, in fact, for the calculation of time correlation functions it is crucial to know how to {unambiguously} define the correlation {function} when a molecule leaves the system and enters in the reservoir. In Ref.\cite{njp} it is discussed how a {precise} definition of the correlation {function} may be a natural consequence of the (first principle) definition of Liouville operator for open systems, {given some well defined properties of the reservoir}; in this paper we developed the formalism of Ref.\cite{njp} further. In general, all modern multiscale techniques dealing with open boundaries (see e.g.\cite{matepj,prx,h-ad,pande,ensing,heyd}) need a clear formulation ({extension})of Liouville's theorem (and related operator) for varying $N$ in order to justify any statistical sampling/averaging performed over the produced trajectories. In this perspective, the aim of this paper is not that of providing a final solution to the problem, but actually is that of laying the basis of a discussion starting from an analysis of what can be concluded according to the research available today. 

\section{{Open quantum systems and reduction to the classical case}}
The study of open quantum systems is a subject of high interest in modern physics and thus the associated physical concepts and related formalism have been extensively treated so that the formal backbone of the theory is very solid \cite{petruc}. In particular the  paper of G.Lindblad \cite{linda} is of central importance; in this paper, about generators of quantum dynamical semigroups, a general form of a certain class of Markovian quantum mechanical  master equations is derived. This work can also be used to describe classical systems in a grand ensemble as well.
The starting point is the equation for the time evolution of the density matrix $\rho(t)$:
\begin{equation}
\dot{\rho(t)}=L(\rho)=-i[H,\rho]+\frac{1}{2}\sum_{j}([L_{j}\rho,L^{+}_{j}]+[L_{j},\rho L^{+}_{j}])
\label{lineq}
\end{equation}
where $H$ is the Hamiltonian, $L_{j}$,$L_{j}^{+}$ are operators which describe the interaction of the system with a reservoir; they are called Lindblad operators while equation \ref{lineq} is also called Kossakowski-Lindblad equation \cite{kosslinda}. The Kossakowski-Lindblad equation describes the most general case of quantum (non linear) evolution of a system embedded in a certain environment. Due to the term $\sum_{j}([L_{j}\rho,L^{+}_{j}]+[L_{j},\rho L^{+}_{j}])$, Eq.\ref{lineq} has the form of a rate equation (quantum jumps in the state of the system due to the action of the external environment) where $[L_{j}\rho,L^{+}_{j}]$ and $[L_{j},\rho L^{+}_{j}]$ can be interpreted as transition rates between two events {(see also note \cite{note})}.
 Under the condition of flux balance: $\sum_{j}([L_{j}\rho,L^{+}_{j}]+[L_{j},\rho L^{+}_{j}])=0$, $\rho(t)$ the stationary solution for $\rho(t)$,  in case of a thermal bath (heat reservoir), is the density matrix of a canonical ensemble. The mathematical analysis of such concepts has been extensively done in Ref.\cite{linda} but, for our current focus, there is one important concept that we can transfer into the treatment of classical systems in a grand ensemble: the Liouville operator in presence of a reservoir takes the form given by Lindblad and the specific action of the reservoir must be well defined through the definition of Lindblad operators. For classical systems $\rho$ is the probability distribution (equivalent of the density matrix of quantum systems) and it is defined as, $\rho(X_{N},N,t)$, where $X_{N}$ is a point in the phase space, $N$ the total number of particles; moreover the commutator $[*,*]$ becomes the Poisson bracket $\{*,*\}$. The classical equivalent of Eq.\ref{lineq} is the standard Liouville equation, plus the corresponding classical term of the Lindblad operators. This latter depends on the specific definition of the reservoir and thus it is model-dependent. Below we treat one specific model of reservoir which is general enough to be of relevance in molecular simulation studies in a grand ensemble.
\subsection{Bergmann-Lebowitz model of reservoir}
Bergmann and Lebowitz (BL) \cite{leb1,leb2} (and subsequently Lebowitz and Shimony \cite{leb3}) proposed a generalization of Liouville's equation to systems that can exchange matter with a reservoir. This work appeared much before the publication of Lindblad's paper, however the BL model can be seen, from the formal point of view, as a specific case of the general approach of Ref.\cite{linda}. 
The model is based on the physical principle that each interaction between the system and the reservoir is characterized by a discontinuous transition of a system from a state with $N$ particles ($X^{'}_{N}$) to one with $M$ particles ($X_{M}$). Importantly, the macroscopic state of the reservoir is not changed by the interaction with the system and thus its microscopic degrees of freedom are not considered {(see also note \cite{notebl})}. The transitions from one state to another are governed by a contingent probability $K_{NM}(X^{'}_{N},X_{M})dX^{'}dt$. The kernel $K_{NM}(X^{'}_{N},X_{M})$ is a stochastic function independent of time and $K_{NM}(X^{'}_{N},X_{M})$ is defined as the probability per unit time that the system at $X_{M}$ has a transition to $X^{'}_{N}$ as a result of the interaction with the reservoir. The term $\sum_{N=0}^{\infty}\int dX^{'}_{N}[K_{MN}(X_{M},X^{'}_{N})\rho(X^{'}_{N},N,t)-K_{NM}(X^{'}_{N},X_{M})\rho(X_{M},M,t)]$ expresses the total interaction between the system and the reservoir and its action is the equivalent of the action of Lindblad operators, thus the general equation of time evolution of the probability is:
\begin{eqnarray}
\frac{\partial\rho(X_{M},M,t)}{\partial t}=-\{\rho(X_{M},M,t),H(X_{M})\}+\\ \nonumber
+\sum_{N=0}^{\infty}\int dX^{'}_{N}[K_{MN}(X_{M},X^{'}_{N})\rho(X^{'}_{N},N,t)-K_{NM}(X^{'}_{N},X_{M})\rho(X_{M},M,t)]
\label{liouvext}
\end{eqnarray}
$H(X_{M})$ is the Hamiltonian of the system corresponding to the point $X_{M}$ and $\{*,*\}$ are the standard Poisson brackets.\\
According to Eq.\ref{liouvext}, if one considers the number of particle as a stochastic variable not explicitly depending on time, the corresponding (classical) Kossakowski-Lindblad equation  or the generalized Liouville's equation can be expressed as:
\begin{equation}
\frac{d \rho(X_{M},M,t)}{dt}=f(X_{M},t)-\hat{Q}\rho(X_{M},M,t)
\label{liouvext2}
\end{equation}
where $f(X_{M},t)=\sum_{N=0}^{\infty}\int dX^{'}_{N}[K_{MN}(X_{M},X^{'}_{N})\rho(X^{'}_{N},t)]$ and $\hat{Q}(*)=\sum_{N=0}^{\infty}\int dX^{'}_{N}[K_{NM}(X^{'}_{N},X_{M}),*]$.\\
If the condition of detailed balance is satisfied: 
\begin{equation}
\sum_{N=0}^{\infty}\int [e^{-\beta H(X^{'}_{N}){ + \beta \mu N}}K_{MN}(X_{M},X^{'}_{N})-K_{NM}(X^{'}_{N},X_{M})e^{-\beta H(X_{M}) { + \beta \mu M}}]dX^{'}_{N}=0
\label{gc-cond}
\end{equation}
it follows that the stationary Grand Ensemble is the Grand Canonical ensemble with density: $\rho_{M}(X_{M},M)=\frac{1}{Q}e^{-\beta H_{M}(X_{M})+ {\beta}\mu M}$
where $\beta=kT$ and $\mu$ the chemical potential.
This is a necessary and sufficient condition for stationarity with respect to the Grand Canonical distribution \cite{leb2,leb3}. In the next section we explore the consequences of such results for the formulation of Liouville's theorem and the definition of Liouville's operator.
\section{{Extension of Liouville's theorem and Liouville's equation to the case of varying $N$}}
\subsection{Liouville's Theorem}
Liouville's  theorem and the corresponding equation are key notions of statistical mechanics. The theorem expresses the concept that a dynamical system composed of $N$ particles conserves its distribution, $\rho$, of positions and momenta (${\bf q},{\bf p}$) along the trajectory.
This concept leads to the equation:
\begin{equation}
\frac{d\rho({\bf q},{\bf p},t)}{dt}=\frac{\partial\rho({\bf q},{\bf p},t)}{\partial t}+\sum_{i=1}^{3N}\left(\frac{\partial\rho}{\partial q_{i}} \dot{q}_{i}+\frac{\partial\rho}{\partial p_{i}} \dot{p}_{i}\right)=0.
\label{liou1}
\end{equation}
One possible (but not unique) solution of Eq.\ref{liou1} is the canonical distribution: $\frac{e^{-\frac{H}{kT}}}{Z}; Z=\int e^{-\frac{H}{kT}}d\Gamma$ with $H$ the Hamiltonian of the system and $\Gamma$ the available phase space.
Often, in mathematical language, Liouville's theorem is formulated as follows: {\it the Lebesgue measure is preserved under the dynamics}.\\
So far we have considered closed systems where $N$ is constant, however what happens in systems (in equilibrium) which exchange particles with external sources?\\
Let us analyze the concept of conservation of Lebesgue measure, that is let us consider an equivalent formulation of Liouville's theorem:
\begin{equation}
\rho({\bf q}_{0},{\bf p}_{0},0)d{\bf q}_{0}d{\bf p}_{0}=\rho({\bf q}_{\tau},{\bf p}_{\tau},\tau)d{\bf q}_{\tau}d{\bf p}_{\tau}.
\label{equivalent}
\end{equation} 
Here ${\bf q}_{0}={\bf q}(0)$, that is ${\bf q}(t)$ at $t=0$; ${\bf p}_{0}$ is defined analogously for the momenta and the same applies to {${\bf q}(t)$ and ${\bf p}(t)$ with  $t=\tau$, moreover we have $\rho({\bf q}_{0},{\bf p}_{0},0)=\rho({\bf q}_{\tau},{\bf p}_{\tau}, \tau)$.}
 We end up in a compact formulation of Liouville's theorem: 
\begin{equation}
d{\bf q}_{0}d{\bf p}_{0}=d{\bf q}_{\tau}d{\bf p}_{\tau};~~~~\forall \tau
\label{ded}
\end{equation}
{Eq.\ref{ded} is a simple consequence of the fact that Hamiltonian dynamics
is a canonical transformation; in Ref.\cite{tuck} it is explained how this relation can be adapted for non-Hamiltonian dynamics.}
In standard textbooks of statistical mechanics and molecular simulation, it is stated that the formalization of Eq.\ref{ded}, is crucial for justifying the fact that ensemble averages can be performed at any point (see e.g. Ref.\cite{tuck}); this is a key concept in molecular simulation.\\
However, the derivation of Eq.\ref{ded} in based on the fact that ${\bf q}_{0}{\bf p}_{0}$ are related to ${\bf q}_{\tau},{\bf p}_{\tau}$ by a coordinate transformation regulated by a Jacobian:
\begin{equation}
J({\bf q}_{\tau},{\bf p}_{\tau},{\bf q}_{0},{\bf p}_{0})=det(Q)
\label{q1}
\end{equation}
where $Q$ is a $6N\times6N$  matrix for a system of $N$ particles defined as:
\begin{equation}
Q_{ij}=\frac{\partial x^{i}_{\tau}}{\partial x^{j}_{0}}
\label{q2}
\end{equation}
where $x_{0}=({\bf q}_{1}(0).....{\bf q}_{N}(0), {\bf p}_{1}(0).....{\bf p}_{N}(0))$ and equivalently $x_{\tau}=({\bf q}_{1}(\tau).....{\bf q}_{N}(\tau), {\bf p}_{1}(\tau).....{\bf p}_{N}(\tau))$. The indices $i,j$ label each of the $6N$ coordinates of $x_{0}$ and $x_{\tau}$, that is: $x^{i}=x^{1}....x^{6N}$ (equivalently for $x^{j}$, with $(x^{1},x^{2},x^{3})=(q_{1}^{x},q_{1}^{y},q_{1}^{z})$ and $(x^{3N+1},x^{3N+2},x^{3N+3})=(p_{1}^{x},p_{1}^{y},p_{1}^{z})$ for example).\\
However in a system where $N$ is variable $det(Q)$ cannot be calculated, since as the system evolves in time the set $x_{0}$ and  $x_{\tau}$ do not necessarily have the same dimension.\\
At this point a natural question arises: on the basis of the results discussed in the first part of the paper, is there a generalized principle, similar to the Liouville's equation for fixed $N$, which extends the concept of Eq.\ref{ded} to the case of variable $N$?\\
As discussed before, if Eq.\ref{gc-cond} is satisfied one would have:
\begin{equation}
\frac{d \rho(X_{M},M,t)}{dt}=0
\end{equation}
corresponding to:
\begin{equation}
\frac{\partial\rho(X_{M},M,t)}{\partial t}=-\{\rho(X_{M},M,t), H(X_{M})\}
\label{lioufin}
\end{equation}
 this equation is formally equivalent to the standard Liouville's equation with fixed number of particles $M$, however this time $\rho(X_{M},M,t)$ and $H(X_{M})$ are instantaneously defined (w.r.t. $M$); similar considerations can be done regarding the definition of Liouville's operator (as it will be discussed later).\\
On the basis of such considerations, a generalized Liouville's theorem extending the formulation for fixed $N$ to the case of a system in contact with a reservoir of particles, may be written in the following form: {\it For systems in contact with a reservoir  of particles, under the condition of statistical flux balance, the Lebesgue measure is conserved for each individual $M$}. {Here we implicitly make the conjecture that the Lebesgue measure cannot be defined globally, but only for single subsets of the phase space, each characterized by a specific number of particles, $M$. This hypothesis is based on the fact that a straightforward definition of a global Lebesgue measure is not obvious.  The fact that $N$ is discrete and a change in $N$ implies a discrete change of the phase space dimensionality, represents a major obstacle. However I do not exclude that it may be possible to define a generalized space where somehow an invariant measure may be defined; this could represent an interesting research program. Here, we have instead proposed a simpler approach, a local definition, meaning for ``local'' a definition of a Lebesgue measure at a given $M$, that is on Canonical hyperplanes.} 
The implication of the above statement would be that Eq.\ref{ded}, may now be extended as:
\begin{equation}
d^{N}{\bf q}_{0}d^{N}{\bf p}_{0}=d^{M}{\bf q}_{\tau}d^{M}{\bf p}_{\tau};~~~~\forall \tau~~~where~~~ M=N.
\label{ded2}
\end{equation}
This approach would be equivalent to the formulation of the problem in terms of canonical hyperplanes as suggested by Peters \cite{peters}; {this means that the condition applies when after some time $\tau$ along a trajectory one returns to the same number of molecules $N$ from which the observation has started. In other terms, in Molecular Dynamics one should sort out instantaneous configurations of a trajectory characterized by the same number of molecules $N$ and for each $N$ then apply the standard Liouville theorem.} 
\section{Practical implications for Molecular Simulation: calculation of equilibrium time correlation functions}
In this section we analyze the consequences of the results of the previous section for quantities of key importance in the physical description of any system: equilibrium time correlation functions. 
The general definition of the equilibrium time correlation function {(e.g. in an NVT ensemble)}, $C_{AB}(t)$ between $A$ and $B$ (physical observables) is\cite{tuck}:
\begin{equation}
\begin{aligned}  
C_{AB}(t) =\langle a(0)b(t)\rangle &  =\int d{\bf p}d{\bf q}f({\bf p},{\bf q}) a({\bf p},{\bf q})e^{iL{t}}b({\bf p},{\bf q})\\
& =\int d{\bf p}d{\bf q}f({\bf p},{\bf q}) a({\bf p},{\bf q})b({\bf p}_{t}({\bf p},{\bf q}),{\bf q}_{t}({\bf p},{\bf q}))
\end{aligned}
\label{eq1}
\end{equation}
$a({\bf p},{\bf q})$ and $b({\bf p},{\bf q})$ are functions in phase space which correspond to $A$ and $B$ (respectively), $f({\bf p},{\bf q})$ is the equilibrium distribution function and $iL$ is the Liouville operator. The general notation is the same used in the guiding reference Ref.\cite{tuck} and ${\bf p}_{t}({\bf p},{\bf q}),{\bf q}_{t}({\bf p},{\bf q})$ indicates the time evolution at time $t$ of the momenta and positions with ${\bf p},{\bf q}$ initial condition. For a system at fixed $N$ (canonical ensemble), Eq.\ref{eq1} is written as:
\begin{equation}
C_{AB}(t)=\frac{1}{Q_{N}}\int d{\bf p}d{\bf q} e^{-\frac{H_{N}({\bf p},{\bf q})}{kT}}a({\bf p},{\bf q}) b({\bf p}_{t}({\bf p},{\bf q}),{\bf q}_{t}({\bf p},{\bf q})).
\label{eq2}
\end{equation}
where $Q_{N}$ is the Canonical partition function and $H_{N}({\bf p},{\bf q})$ the Hamiltonian of a system with $N$ (constant) molecules.
It follows that the numerical calculation of $C_{AB}(t)$ is done by calculating $a({\bf p},{\bf q})$ and $b({\bf p}_{t}({\bf p},{\bf q}),{\bf q}_{t}({\bf p},{\bf q}))$ along MD trajectories and then taking the average.
The dynamics generated by Liouville's operator is well defined, since the Hamiltonian of $N$ molecules is well defined at any time $t$:
\begin{equation}
iL=\sum_{j=1}^{N}\left[\frac{\partial H}{\partial{\bf p}_{j}}\frac{\partial}{{\partial{\bf q}^{j}}}-\frac{\partial H}{\partial{\bf q}^{j}}\frac{\partial}{{\partial{\bf p}_{j}}}\right]=\left\{*,H\right\}
\label{lioufix}
\end{equation}
{What does it happen in case of a Grand Canonical ($\mu$ V, T) ensemble?}\\
Let us generalize the formalism of Eq.\ref{lioufix}:
\begin{equation}
C_{AB}(t)=\frac{1}{Q_{GC}}\sum_{N}\int d{\bf p}_{N}d{\bf q}_{N} e^{-\frac{[H_{N}({\bf p}_{N},{\bf q}_{N})-\mu N]}{kT}}a({\bf p}_{N},{\bf q}_{N}) b({\bf p}_{t}({\bf p}_{N},{\bf q}_{N}),{\bf q}_{t}({\bf p}_{N},{\bf q}_{N})).
\label{eq3}
\end{equation}
where $Q_{GC}$ is the Grand-Canonical Partition function, $\mu$ the chemical potential and $N$ the number of particles (now varying in time) of the system.
The question now is about how to interpret the quantity $b({\bf p}_{t}({\bf p}_{N},{\bf q}_{N}),{\bf q}_{t}({\bf p}_{N},{\bf q}_{N}))$; at a given time $t$ the system evolved from its initial condition and may have a number of particles $N^{'}$ different from the initial state.
The BL model (see previous discussions and Refs.\cite{leb2,leb3,njp}) allows to make sense of $b({\bf p}_{t}({\bf p}_{N},{\bf q}_{N}),{\bf q}_{t}({\bf p}_{N},{\bf q}_{N}))$ in numerical simulations. 
In fact in Eq.\ref{eq1}, when extended to the case of an open system, the propagator {$e^{iL{t}}$} must be substituted by the propagator characterized by the extended Liouville operator which includes the action of the reservoir via the (additional) Lindblad operator.
Let us write the extended Liouville operator as $L^{ext}=L+L^{Lindblad}$, thus its action on $b({\bf p},{\bf q})$ can be written as :${e^{i(L+L^{Lindblad})t}}b({\bf p},{\bf q})=b({\bf p}_{t}({\bf p}_{N},{\bf q}_{N}),{\bf q}_{t}({\bf p}_{N},{\bf q}_{N}))$. {In order to emphasize the problem we want to discuss, let us consider time correlation functions based on single-molecule properties, such as, for example, molecular velocity-velocity time autocorrelation functions or molecular dipole-dipole autocorrelation function.  The definition of the velocity  autocorrelation function is:
\begin{equation}
C_{VV}(t) = \frac{1}{N}\sum_{i=1}^{N}\frac{\langle v_{i}(t) \cdot v_{i}(0) \rangle}{\langle v_{i}(0) \cdot v_{i}(0) \rangle}
\end{equation}
where $\langle \cdot \rangle$ denotes the ensemble average and $\langle v_{i}(t) \cdot v_{i}(0) \rangle$ computes 
the correlation between the velocities of $i^{th}$ molecule at initial time 0 and at a time $t$. In the same way one may define the dipole auto correlation function: $C_{\mu\mu}(t) = \frac{1}{N}\sum_{i=1}^{N}\frac{\langle \mu_{i}(t) \cdot \mu_{i}(0) \rangle}{\langle \mu_{i}(0) \cdot \mu_{i}(0) \rangle}$ or other time correlation functions based on the properties of single molecules.
In molecular dynamics the correlation functions introduced above require the calculation, in a time window $\Delta t$, of the value of $C_{AB}(t)$ for each molecule. Such a procedure, in an open system, requires the unambiguous and precise definition of ``{\it each molecule}''.}\\
In fact, differently from the case at fixed $N$ where ``{\it each molecule}'' is trivially defined, for open systems we would have to consider three cases:
\begin{itemize}
\item (i) Molecules which remains in the system within the time window considered
\item (ii) Molecules which initially are in the system and then enters into the reservoir  within the time window considered
\item (iii) Molecules which entered, within the time window considered, from the reservoir into the system
\end{itemize}
{For the case (i) we have  that the Lindblad operator applied to a molecule $i$ is such that ${e^{i(L^{Lindblad})t}}b_{i}({\bf p}_{i},{\bf q}_{i})=b_{i}({\bf p}_{i},{\bf q}_{i})$,because in this case the Lindblad operator ({by definition of its action through $K_{MN}$ and $K_{NM}$ in the Bergmann-Lebowitz model, as discussed before}) does not act directly on molecule $i$, that is, in terms of physics, the reservoir does not  directly modify the microscopic status of molecule $i$, since the molecule remained in the system. As a consequence in the definition of $C_{AB}(t)$ only Liouville's operator (as for the case at fixed $N$) is involved. For the case (ii) instead we have that the action ${e^{i(L^{Lindblad})t}}b_{i}({\bf p}_{i},{\bf q}_{i})$, according to the specific definition of reservoir of Bergmann and Lebowitz (i.e. according to the action of $K_{MN}$ and $K_{NM}$), is similar to the action of an annihilation operator which annihilates the microscopic identity of molecule $i$ once it enters into the reservoir. As a consequence it removes its contribution to the correlation function by destroying the quantity $b_{i}({\bf p}_{i},{\bf q}_{i})$ corresponding to the specific molecule since the index $i$ of such molecule does not exist anymore} (see also discussion in the Appendix). {It must be noticed that the fact that the correlation does not exist because the molecule, once it enters into the reservoir, does not exist anymore it is not equivalent to say that the correlation becomes zero. This is not true from the physical point of view, instead the molecule simply does not contribute to the average of the correlation function because,  by definition, does not posses a correlation. Finally the case (iii) is trivial because $a_{j}({\bf p}_{j},{\bf q}_{j})$ of the molecule $j$, not present in the system before, is not defined; that is a molecule entering  from the reservoir, may posses instantaneous microscopic memory once it enters in the system (i.e. $b_{j}({\bf p}_{j},{\bf q}_{j})$ is defined), but do not have memory of preceding time (i.e. $a_{j}({\bf p}_{j},{\bf q}_{j})$) , thus by definition the integral in the calculation of $C_{AB}(t)$ is not done over particles that are not present in the system at the initial time.} The important point of this discussion is that once the action of the reservoir is specified, then it follows an unambiguous ``{\it numerical}'' recipe on how to count molecules for $C_{AB}(t)$ in a molecular dynamics study. For the case of Bergmann-Lebowitz model, according to the discussion above, the following definition arises:{\it ``When, within the observation time window, a molecule crosses the border of the system and enters in the reservoir, its contribution to the correlation function must be neglected because, given the specific definition of the reservoir, the microscopic identity is deleted''}.\\
An interesting  consequence of the definition above is that a correlation function related to a certain physical process depends on how local in space a certain process is, because of the finite size (usually relatively small) of an open system and thus of the number of particles considered. However, at the same time, because of the lost of microscopic memory when a molecule enters into the reservoir, the correlation function depends also on the locality in time of the physical process. This means that, although the physical meaning is clear, the correlation function calculated in an open system differs from the correlation function calculated in a closed system, and they do coincide only in the thermodynamic limit {(see also note \cite{noteterm} for indirect implications of this result)}.
Moreover, the definition above, although may look natural, is actually not obvious or trivial, in fact, for example, one can also define a subsystem of a large system and consider the embedding larger part as a natural reservoir {(as further methodological and theoretical examples of Grand Canonical simulation set up that cannot be defined as a subsystem of a large microscopic (but closed) system and must follow the treatment proposed here see Refs.\cite{rafa,matejjcp})}. Since the microscopic details of the reservoir in such case are known, then the Lindblad operator may be explicitly defined through the microscopic interaction in the system-reservoir term of the Hamiltonian. In such case the numerical definition for time correlation function of the system is obviously different from the one above (with a different physical interpretation), since the microscopic memory may be considered, in this case, not lost, thus the correlation function has to be calculated also when a molecule leaves the system and enters in the reservoir; this implies that locality in time does not play a role. Both definitions have a physical sense, but such a physical sense becomes unambiguous once the formal action of the reservoir (and thus its corresponding generalized Liouville operator or its Lindblad operator) is specified. It must be noticed that the derivation of an extended Liouville operator does not apply only to the calculation of equilibrium time correlation functions but to the calculation, in open systems, of observable characterized by a response in time. An example is the Onsager-Kubo method developed by Ciccotti and coworkers \cite{cic1,cic2,jctchan}, which, if applied to open systems, requires, similarly to the case of equilibrium time correlation functions, the definition of extended Liouville operator as discussed here; {in this case one shall consider an additional perturbation to the unperturbed propagator, ${e^{i(L+L^{Lindblad})t}}$, which then gives rise to the situations (i), (ii) and (iii) as discussed before}.
\section{Discussion and Conclusions}
We have discussed the problem of the extension of Liouville's theorem for systems with varying number of particles. Starting from the general formalism for open quantum systems we have {proposed}, through the work of Bergmann and Lebowitz, a generalization of Liouville's equation and Liouville's Theorem for classical systems. Finally we have discussed its relevance for molecular simulation studies in a grand ensemble, in particular for the {calculation} of equilibrium time correlation function. A clear formulation of Liouville's theorem Liouville's operator and Liouville's equation is mandatory for justifying the validity of numerical calculations. { In any case, as the premise of the paper states, the intention of this work is to provide only the basis of a discussion about this issue by discussing (some of) its relevant aspects. Regarding concrete examples, we have discussed only one specific theoretical model of reservoir, chosen because in its current formulation is already of practical utility for molecular simulation. However, as reported in the appendix, there are other models which in the current form are not practical for simulation but may represent the basis for designing new reservoirs; our hope is that this paper may inspire further development in the field.}
\section*{Acknowledgment}
I would like to thank Giovanni Ciccotti and Carsten Hartmann for uncountable discussions about the problem of Liouville's Theorem in open systems and Matej Praprotnik for a critical reading of the manuscript and useful suggestions. This research has been funded by Deutsche Forschungsgemeinschaft (DFG) through grant CRC 1114.
\section{Appendix}
{In the discussion above we have qualitatively described the action of the Lindblad operator in the Bergmann-Lebowitz model as similar to the action of an annihilation or creation operator. The question of whether or not one can explicitly write the action of the reservoir of Bergmann and Lebowitz in terms of annihilation or creation operators may be, within the framework of molecular simulation, a question of practical convenience. For example, in another seminal paper about open systems, Emch and Sewell \cite{emsew} derive a formally elegant Liouville-like equation based on the explicit use of well defined operators. The interesting part, related to the annihilation or creation operator, is in the definition of interaction Hamiltonian term between the system and the reservoir (equivalent to the integral term of the Bergmann-Lebowitz model).
They define such term as:
\begin{equation}
H_{I}=\int_{\Omega_{R}}dx\int_{\Omega_{S}}dy V(x,y)J_{R}(x)J_{S}(y)
\label{intes}
\end{equation}
where $\Omega_{R}$ and $\Omega_{S}$ are the phase space of the reservoir and of the system respectively, $V(x,y)$ is the interaction potential between the reservoir and the system and $J_{R}(x)$, $J_{S}(y)$ are operators acting respectively on the $x$ space of the reservoir and on the $y$ space of the system. Next the authors state: {\it ``More generally $J_{R}$ and $J_{S}$ might be function of the creation and annihilation operators for particles in the reservoir and in the system''}. One shall consider that  in such specific case the authors are thinking of a quantum system, however, formally, the problem is equivalent to that of a classical system (as considered by us). As stated before, for molecular simulation the formulation of Emch and Sewell, as it is currently done, would not be practical for one main reason, that is the requirement of an explicit definition of $V(x,y)$. In fact the model of Emch and Sewell is based on the idea of projector operators where the underlying microscopic character of the reservoir is projected out, however it implies that the reservoir has an explicit microscopic evolution}.


\begin{thebibliography}{24}
\bibitem{noteaddress}
In the AdResS method used in Refs.\cite{noi,kk-prot} the regions characterized by different molecular resolution are coupled by a space-dependent interpolation between the atomistic and coarse-grained force acting on the molecules. I have shown that for such a scheme a global Hamiltonian cannot exist and thus the very definition of  a global Liouville operator is not possible (see Ref.\cite{premio}). In such case it is mandatory to define the Liouville operator for the atomistic region where quantities are calculated.
\bibitem{notebl}
{
The Bergman-Lebowitz model considers the most general case of a system in contact with several reservoirs each acting independently of the other; the total action is the sum of each reservoir's action on the system. The case of a single reservoir, as it is done in this work, is then a sub-case of the more general model.
The interaction with this reservoir is impulsive/stochastic in the sense that the system jumps in a discontinuous way regarding the number of particles and the new particles entering into the system have a velocity consistent with the temperature of the reservoir and thus, in our case, being in thermal equilibrium, of the system. However, an explicit back-reaction from the system to the reservoir that changes the macroscopic (thermodynamic) state of the reservoir is not considered.}
\bibitem{note}
{An explicit expression of the Lindblad operator may have the form: $\frac{1}{2}\sum_{j}([L_{j}\rho,L^{+}_{j}]+[L_{j},\rho L^{+}_{j}])=\sum_{n=1}^{N}\sum_{m=1}^{M}[l_{mn}(a_{n}\rho a^{+}_{m}-\frac{1}{2}(\rho a^{+}_{m}a_{n}+a^{+}_{m}a_{n}\rho))]$, where $l_{mn}$ are transition probabilities from state $n$ to state $m$ and $a_{m,n}$ is a creation annihilation operator which creates or destroy a state $m$ ($n$) in the density matrix. How it is discussed in this paper, the Bergmann- Lebowitz model for classical system can be seen as a realization of such an operator for classical cases where the system is allowed with some rate/probability to go from a number of molecules N to a number of molecules M with the implicit creation (annihilation) of molecules.}
\bibitem{noteterm}
{ From this work it can be concluded that 
calculated time correlation functions will
depend upon both the simulated ensemble and also the dynamical
equations of motion. In particular, even if two dynamics generate the same equilibrium ensemble, they may in principle generate different time-correlation
functions. This aspect has been treated in Ref.\cite{leimkuhler}, where the influence of the thermostat on dynamical properties of a system is extensively treated.}
\end{thebibliography}
\end{document}